\documentclass[aps,prb,twocolumn,floatfix,showpacs]{revtex4}
\usepackage{amsfonts}
\usepackage{graphicx}
\usepackage{dcolumn}
\usepackage{bm}
\usepackage{amsmath}
\usepackage{amssymb}

\begin{document}
\title{Effects of disorder on magnetic vortex gyration}
\author{Hongki Min$^{1,2}$}
\email{hmin@umd.edu}
\altaffiliation[Current address: ]
{Condensed Matter Theory Center, Department of Physics, 
University of Maryland, College Park, Maryland 20742, USA}
\author{R. D. McMichael$^{1}$}
\author{Jacques Miltat$^{1,2,3}$}
\author{M. D. Stiles$^{1}$}
\affiliation{
$^{1}$Center for Nanoscale Science and Technology, National Institute of Standards and Technology, Gaithersburg, Maryland 20899-6202, USA\\  
$^{2}$Maryland NanoCenter, University of Maryland, College Park, Maryland 20742, USA\\  
$^{3}$Laboratoire de Physique des Solides, Universit\'{e} Paris Sud, CNRS, UMR 8502, F-91405 Orsay, France 
}
\date{\today}

\begin{abstract}
A vortex gyrating in a magnetic disk has two regimes of motion in the
presence of disorder.  At large gyration amplitudes, the vortex core
moves quasi-freely through the disorder potential.  As the amplitude
decreases, the core can become pinned at a particular point in the
potential and precess with a significantly increased frequency.  In
the pinned regime, the amplitude of the gyration decreases more
rapidly than it does at larger precession amplitudes in the quasi-free
regime.  In part, this decreased decay time is due to an increase in
the effective damping constant and in part due to geometric distortion
of the vortex.  A simple model with a single pinning potential
illustrates these two contributions.
\end{abstract}

\maketitle
\normalsize


\section{Introduction}

\begin{figure}
\includegraphics[width=1\linewidth]{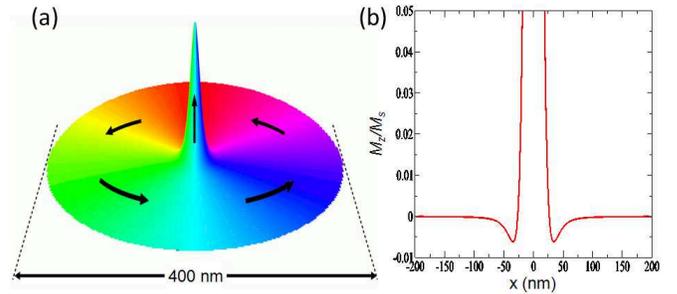}
\caption{(Color online). (a) A typical vortex structure in
  a Ni$_{80}$Fe$_{20}$ disk with 400~nm diameter and 10~nm
  thickness calculated as described in Sec.~\ref{sec:micromag}. The
  color indicates the in-plane angle of the 
  magnetization, and the arrows indicate the approximate magnetization
  direction. (b) Cross section of the $z$-component magnetization
  along the center of the vortex core in (a).}
\label{fig:vortex}
\end{figure}

In disks of magnetic material, the ground-state magnetic configuration
is commonly a magnetic vortex state which forms due to the interplay
between magnetostatic and exchange energies.  In the vortex structure
the magnetization in the wall curls around a vortex core and points
out of the plane at the core region, as illustrated in
Fig.~\ref{fig:vortex}.  Alignment of the magnetization parallel
to the edge of the disk minimizes the magnetostatic energy and the
magnetization pointing out of the plane in the core avoids a singularity in the
exchange energy.  In the vortex state, the excitation spectrum is
significantly modified compared to that of a uniform magnetization.
In particular, there is a low-frequency gyration mode, in which the
vortex core orbits around its minimum energy location.  

The dynamics of a vortex state have been studied experimentally by
time-resolved Kerr
microscopy,\cite{park2003,zhu2005,compton2006,compton2010}
time-resolved scanning transmission x-ray
microscopy,\cite{choe2004,kuepper2007,vansteenkiste2008} and microwave
reflection technique\cite{novosad2005} and theoretically using a
collective coordinate approach or micromagnetic
simulations,\cite{guslienko2002,guslienko2005,guslienko2006} which
show gyration frequencies typically in the subgigahertz range.  If the vortex
structure is excited strongly enough, the vortex core switches
magnetization direction.
Vortex core switching has been observed using time-resolved scanning
transmission x-ray microscopy\cite{waeyenberge2006,vansteenkiste2009a}
and modeled using a collective coordinate approach or micromagnetic
simulations.\cite{guslienko2008,vansteenkiste2009b,apolonio2009,silva2008}
These studies show a reversal of the vortex core magnetization with a
relatively low threshold magnetic field on the order of milliteslas by an
in-plane oscillating external magnetic field.  Recent experiments have
also studied vortex gyration and core reversal under excitation by
current in the plane of the
disk.\cite{vansteenkiste2009a,kasai2006,yamada2007,hertel2007,pribiag2007,pribiag2009}
Excitation by current perpendicular to the disk has bee studied
theoretically.\cite{caputo2007,sheka2007,gaididei2010,guslienko2010}
Current induced motion is beyond the scope of the present article.

Of particular interest here are experiments that measure the dynamics
of vortices in disordered samples,\cite{uhlig2005} particularly
the recent experiments by Compton \textit{et al.},\cite{compton2006,compton2010} who measured vortex gyration in the 
presence of disorder.  In these experiments, the vortex core is
displaced by a static in-plane magnetic field and then a magnetic
field pulse is applied to excite the vortex motion. At small 
magnetic-field-pulse amplitudes, the vortex gyrates about its equilibrium
position with a frequency that is characteristic of the local disorder
potential.  At large pulse amplitudes, the vortex gyrates with a
frequency determined by the magnetostatic energy of the disk. Between
the two amplitudes, sharp transitions in the gyration frequency
correspond to pinning or depinning of the vortex at local defects.

The collective coordinate approach mentioned above readily explains
many aspects of these experiments.\cite{compton2006,compton2010} Here,
we address an aspect that has not yet been considered; that is, whether
disorder changes the effective damping constant needed for a
description of the behavior in terms of collective coordinates.  This
approach is motivated by our recent theoretical demonstration
\cite{min2010} that when a magnetic domain wall propagates along a
magnetic nanowire in the presence of disorder, the effective damping
is enhanced as disorder increases, leading to increased or decreased
domain wall velocity depending on the conditions. As a domain wall
moves through disorder, internal degrees of freedom get excited,
increasing the energy dissipation rate and thus the effective
damping.  The results of Compton \textit{et al.} provide much more detailed
information about the interaction of vortices with disorder than is
typically accessible in experiments on domain wall motion in
nanowires.  This detail suggests that it might be possible to
quantitatively connect theory and experiment relating enhanced
damping. 

In this article, we describe micromagnetic simulations of vortex
gyration both in the complex case of random disorder and also in the
simple case of a single pinning potential.
Our results indicate that disorder, which exists inevitably in real
experiments, affects the vortex dynamics in a way that can be
interpreted as an enhancement of the effective damping. 
In Sec.~\ref{sec:method} we describe the theoretical approach, in
particular the micromagnetic simulations and the description of the
dynamics in terms of a reduced set of degrees of freedom.  In
Sec.~\ref{sec:VGdisorder} we compare our simulations with the
results of Ref.~\onlinecite{compton2006} and Ref.~\onlinecite{compton2010}
showing the transition between 
quasi-free gyration and pinned gyration.  In
Sec.~\ref{sec:VGsingle}, we study this transition in a system with
a single pinning potential to make clear the origin of various
effects.  Finally, in Sec.~\ref{sec:discussion}, we discuss the
implications of these results.

\section{Method}
\label{sec:method}
\subsection{Collective coordinate approach}
Magnetization dynamics in a magnetic field can be described by the
Landau-Lifshitz-Gilbert equation
\begin{equation}
\label{eq:LLG}
\dot{\bf M}=\gamma {\bf H}_{\rm eff}\times {\bf
  M}+\frac{\alpha_{\rm G}}{M_{\rm s}}{\bf M}\times\dot{\bf M},
\end{equation}
where ${\bf H}_{\rm eff}$ is the effective magnetic field including
the external, exchange, demagnetization, and anisotropy fields;
$\gamma$ is the gyromagnetic ratio; $M_{\rm s}$ is the saturation
magnetization; and $\alpha_{\rm G}$ is
the Gilbert damping constant.  In the calculations described below, we
study the dynamics of vortex gyration with Eq.~(\ref{eq:LLG}) using a
fixed value of $\alpha_{\rm G}$.  We discuss an effective damping parameter
$\alpha$ in the context of a description of the motion in
terms of collective coordinates, which we describe next.

Vortex motion is frequently studied in models that adopt
a description of vortex structures in terms of a limited
number of collective
coordinates.\cite{thiele1973,thiaville2005,he2006,tretiakov2008,clarke2008}
For 
the simplest approximation, the vortex gyration is described by a
two-dimensional vector ${\bf X}=(X_1,X_2)$ describing the vortex core
position in a plane. Then Eq.~(\ref{eq:LLG}) reduces to
\cite{thiele1973,thiaville2005,he2006,tretiakov2008,clarke2008}
\begin{equation}
\label{eq:collective_VW2}
\alpha{\bf D}\dot{{\bf X}}={\bf F}+\dot{{\bf X}}\times{\bf G},
\end{equation}
where
\begin{eqnarray}
\label{eq:D_F_G}
D_{ij}&=&\frac{1}{M_{\rm s}^2}\int dV\, \frac{\partial{\bf M} }{ \partial X_i}\cdot
\frac{\partial{\bf M} }{ \partial X_j}, \\ 
{\bf F}&=&\frac{\gamma}{M_{\rm s}} \int dV\,
{\bf H}_{\rm eff}\cdot \frac{\partial {\bf M}}{ \partial
  {\bf X}}=-\frac{\gamma }{ \mu_0 M_{\rm s}} \frac{\partial E }{ \partial {\bf X}},
\nonumber \\ 
{\bf G}&=&{\hat{\bf z}}\frac{1}{M_{\rm s}^3}\int dV\, {\bf M}\cdot
\left(\frac{\partial{\bf M} }{ \partial X_1} \times
     \frac{\partial{\bf M} }{ \partial X_2}\right), \nonumber
\end{eqnarray}
and $E$ is a total energy functional whose derivative gives
$\mu_0{\bf H}_{\rm eff}\equiv-\frac{\delta E}{ \delta {\bf M}}$. Note that
$\alpha{\bf D}$ is a symmetric matrix which characterizes viscous
friction, ${\bf F}$ is a generalized force, and ${\bf G}$ is a
gyrotropic tensor ($\dot{{\bf X}}\times{\bf G}$ is the gyrotropic
force) which characterizes magnetization precession. Thus the dynamic
properties of the vortex state are similar to that of a two-dimensional
massless charged particle moving through a medium with a viscosity
tensor $\alpha{\bf D}$ in the presence of an in-plane electric field
${\bf F}$ and a perpendicular magnetic field ${\bf
  G}$.\cite{clarke2008} The ``masslessness'' of the dynamics is
inherited from Eq.~(\ref{eq:LLG}).  
In the absence of disorder, such a dynamics can be solved analytically by assuming harmonic oscillation 
for the vortex core. \cite{guslienko2002,guslienko2005,guslienko2006,wysin2010}

Assume that $E$ has a quadratic dependence on the radial position of
the vortex core $r$ so that
${\bf F}=-k r {\bf \hat{r}}$, where $k$ is a constant which characterizes
the shape of the potential in which the vortex core gyrates.  In polar
coordinates, Eq.~(\ref{eq:collective_VW2}) can be expressed in the
following matrix form
\begin{equation}
\label{eq:collective_VW2_matrix}
\alpha\left(
\begin{array}{cc}
D_{rr}     &D_{r\phi}     \\
D_{r\phi}  &D_{\phi\phi}  \\
\end{array}
\right)
\left(
\begin{array}{c}
\dot{r}\\
r\dot{\phi}\\
\end{array}
\right)
=
\left(
\begin{array}{c}
-kr+G r\dot{\phi} \\
-G \dot{r} \\
\end{array}
\right),
\end{equation}
where $\phi$ is the azimuthal angle of the vortex core position from
the gyration center.

Assuming $r=r_0 \exp(-t/\tau)$ and $\phi=2\pi f t$, and by eliminating $k$, we have 
\begin{equation}
\label{eq:damping_relation}
2\pi f\tau C \alpha=1,
\end{equation}
where $C=D_{\phi\phi}/(G+\alpha D_{R\phi})$.  These four parameters,
the gyration frequency $f$, the decay time $\tau$, the effective damping
parameter $\alpha$, and the geometrical factor $C$, are the focus of
the subsequent analysis. A related analysis was carried out by Compton \textit{et al.}\cite{compton2006,compton2010} 
to model behavior of the precession frequency
$f$ in the presence of a pinning potential.  They assumed that $D$ and
$G$ were constant.  Here, we focus on the decay time, $\tau$, which is
also measurable and examine the extent it is modified by changes in
$D$, $G$, and the effective damping, $\alpha$.
The geometrical factor $C$ is related to the deformation
of a vortex structure, which is hard to measure but can be
evaluated in a simulation using Eq.~(\ref{eq:D_F_G}).  In analyzing
measurements of the decay time, it would be tempting to assume that
$C$ is constant and 
ascribe the observed changes in $f\tau$ as due to changes in
$\alpha$.  Here, we test the degree to which that would be correct.
Note that Eq.~(\ref{eq:damping_relation}) is explicitly independent of $k$, thus
independent of a specific potential shape in which a vortex
gyrates.  However, each of the parameters in
Eq.~(\ref{eq:damping_relation}) depends on $k$.  In particular $C$
depends on the vortex shape, and hence does depend weakly
on $k$ for the disks of interest here.

Now imagine a vortex gyration experiment in a disk with a pinning
potential at the center. When a large enough field pulse is applied, a
vortex gyrates outside of the pinning potential and as the orbit
decays, it is eventually
trapped by the potential.  The frequency of the precession $f$ and
its decay time $\tau$ change when the vortex becomes pinned
\begin{equation}
\label{eq:damping_ratio}
\frac{f_0 \tau_0}{ f \tau}
=
\frac{\alpha_{\rm eff}}{\alpha_0}
\frac{C}{C_0}, 
\end{equation}
where $\alpha_{\rm eff}$, $f$, $\tau$, and $C$ indicate values in the
trapped region while those with the subscript $0$ indicate values in
free region.  The frequency and decay time can be measured so if
the vortex geometry stays the same, that is if $C$ remains constant,
measurements of $f$ and $\tau$ could be used to infer the change in the effective
damping constant. In Sec.~\ref{sec:VGsingle}, however, we will show that 
$C$ is not constant in the presence of disorder.

When the motion of the vortex is well described by the collective
coordinates, we expect that $\alpha_{\rm eff}=\alpha_0=\alpha_{\rm G}$.  
However, if other modes of the system are excited, the total energy dissipation
rate would increase leading to an increase in the effective values of
$\alpha$ and faster decay (smaller $\tau$).  One of the goals of
this work is to characterize this increase and to compare it to the
increase observed in the modeled motion of vortex domain
walls.\cite{min2010}  In the subsequent sections, we use micromagnetic 
simulations to study how the effective damping is changed during the
gyration motion and how it affects the decay time of the gyration.
We stress that the ``real'' damping constant does not change, but the
value consistent with the collective coordinate description does.

\subsection{Micromagnetic simulation}
\label{sec:micromag}

We compute the dynamics of the vortex state through numerical solution of
Eq.~(\ref{eq:LLG}) using the Object Oriented MicroMagnetic Framework
(OOMMF).\cite{oommf} We set up a Ni$_{80}$Fe$_{20}$ disk with a 400~nm
diameter and a 10~nm thickness, as shown in
Fig.~\ref{fig:vortex}(a). We use computational cells that are uniform
through the thickness and have an in-plane
size of 2.5~nm. For material constants, we use the
saturation magnetization $M_{\rm s}$=800 kA/m, damping constant
$\alpha_{\rm G}$=0.01 and either a fixed value of the exchange stiffness constant
$A$=13 pJ/m or the exchange length $l_{\rm ex}=5.7$~nm. 
Note that 
\begin{equation}
l_{\rm ex}\equiv\left[2A/(\mu_0 M_{\rm s}^2)\right]^{1/2}.
\label{eq:lexDef}
\end{equation}

We tested the cell size dependence by studying a system with reduced
size -- 200~nm diameter and 5~nm thickness -- and compared the
results of simulations with cell sizes 1.25~nm and 2.5~nm. We found
that these simulations agree to within 5 \% with no qualitative
differences. Thus, for the bigger system we treat in this
article, with 400~nm diameter and 10~nm thickness, we use a 2.5~nm cell size.

\subsection{Disorder model}

\begin{figure}
\includegraphics[width=1\linewidth]{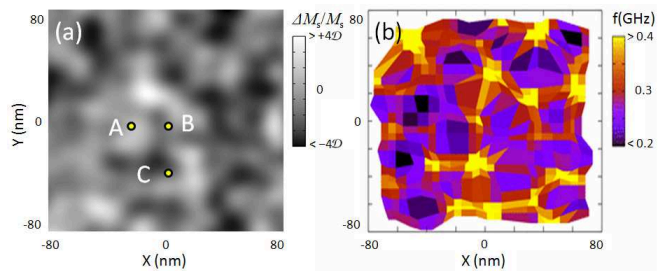}
\caption{(Color online). (a) A typical disorder image with 10~nm
  spatial correlation length, and (b) contour maps of the gyration
  frequency as a function of position. (a) Part of the
  saturation magnetization within a Ni$_{80}$Fe$_{20}$ disk with a 400~nm
  diameter and a 10~nm thickness.  White indicates maxima in the
  magnetization and black minima. These results are based on a model
  of the disorder with ${\cal D}=0.05$ and a fixed exchange stiffness
  constant $A$. (b) The gyration frequency according to
  the color scale on the right.  To move the vortex positions,
  in-plane static magnetic fields from $-10$~mT to $+10$~mT along the
  $x$ and $y$ directions were applied.  The gyration frequency is
  obtained by applying a field pulse of 0.1~mT for 200~ps along the
  $y$ direction.  The final location of the vortex core determines the
  real space position used in the figure.}
\label{fig:disorder_map}
\end{figure}

We motivate our model for disorder on the measurements in
Ref.~\onlinecite{compton2006}.  The article shows magnetic force microscopy
images that reveal thickness fluctuations with a characteristic length
scale of about 10~nm.  Rather than dealing with a full-fledged
model of thickness fluctuations, 
we introduce disorder by varying the
saturation magnetization $M_{\rm s}$ while fixing either $A$ or
$l_{\rm ex}$.  In the main text, we fix $A$ and, in the Appendix, we
discuss the quantitative but not qualitative changes that occur when we
fix $l_{\rm ex}$.  We generate a random, white-noise model for the
variation of $M_{\rm s}$ and convolute it with a Gaussian that has a width
of 10~nm.  A typical disorder image is shown in
Fig.~\ref{fig:disorder_map}(a), showing the smooth variation of $M_{\rm
  s}$ guaranteed by the convolution.  Regions with reduced
magnetization, intended to model thin parts of the sample, are shown
in black and tend to create pinning centers for the vortex core and
locally increase the gyration frequency as shown in Fig.~\ref{fig:disorder_map}(b).

To estimate realistic disorder amplitudes, we compute the variation of
the gyration frequencies to be compared with the measurements in
Ref.~\onlinecite{compton2006}, which show a factor of 2 to 3
variation in resonance frequency as a vortex is scanned over a
disk-shaped sample.  We compute the precession frequency in the limit
of low precession amplitudes as a function of $H_x$ and $H_y$, in
correspondence with the experimental procedure.  In distinction to the
experiment, we can view the equilibrium position of the vortex in real
space and plot the frequency as a function of position to see the
correlations with the disorder image.  The mapping from applied field
to vortex position is responsible 
for the irregular grid seen in Fig.~\ref{fig:disorder_map}(b).
For these calculations, we characterize the size of the disorder by the
ratio of the standard deviation of the fluctuations in the
magnetization to the saturation magnetization, ${\cal
  D}$=$\sqrt{\left<(M({\bf r})-M_{\rm s})^2\right>}/M_{\rm s}$.  We
limit the size of the fluctuations to ensure that the magnetization
stays positive. We find that for a fixed $A$, a disorder value of
${\cal D}=0.05$ gives roughly the same gyration frequency variation as
the experiment. The modeled frequency variation is shown in
Fig.~\ref{fig:disorder_map}(b).

\section{Vortex gyration in a disordered film}
\label{sec:VGdisorder}

\begin{figure}
\includegraphics[width=1\linewidth]{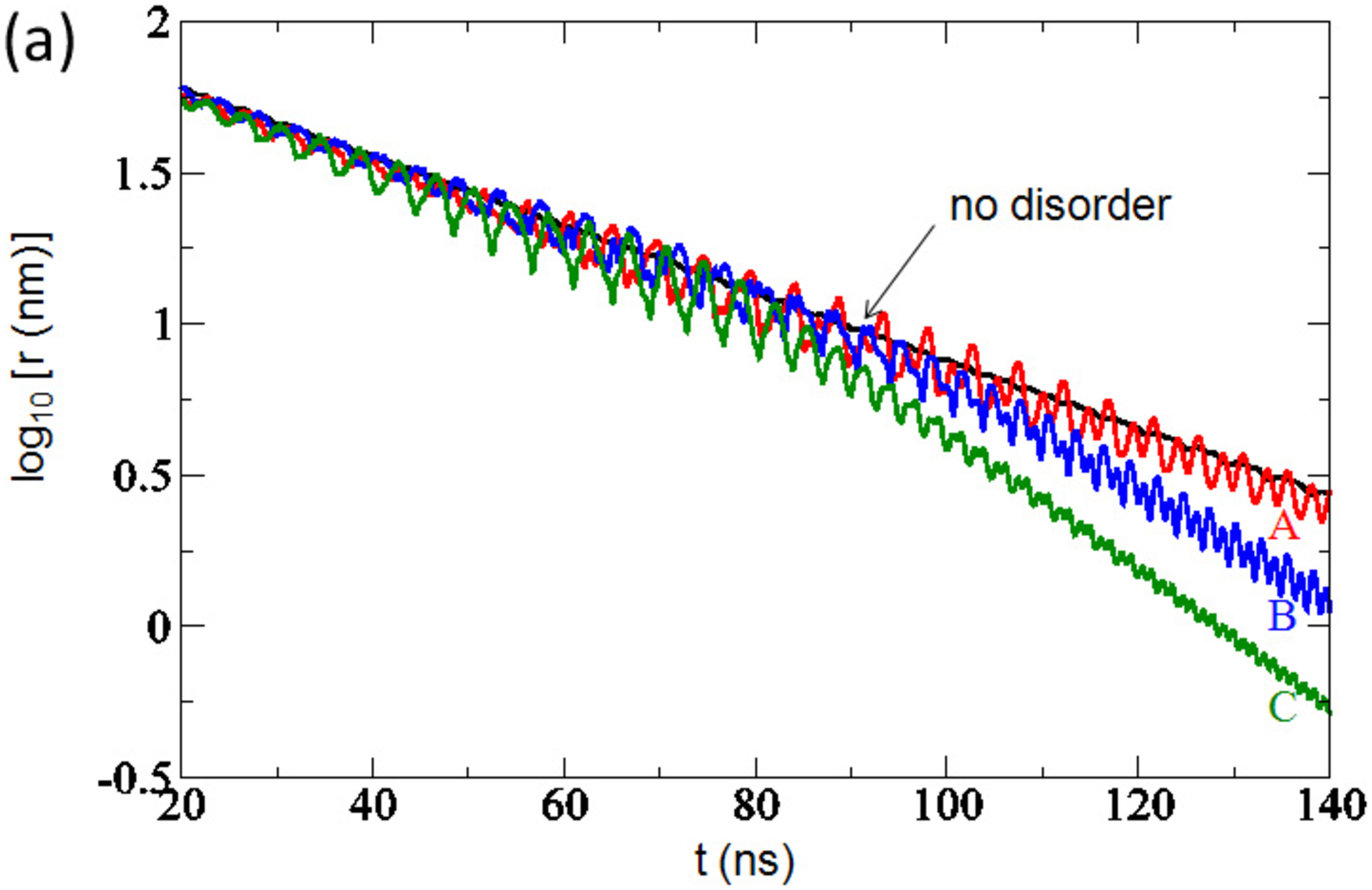}
\includegraphics[width=1\linewidth]{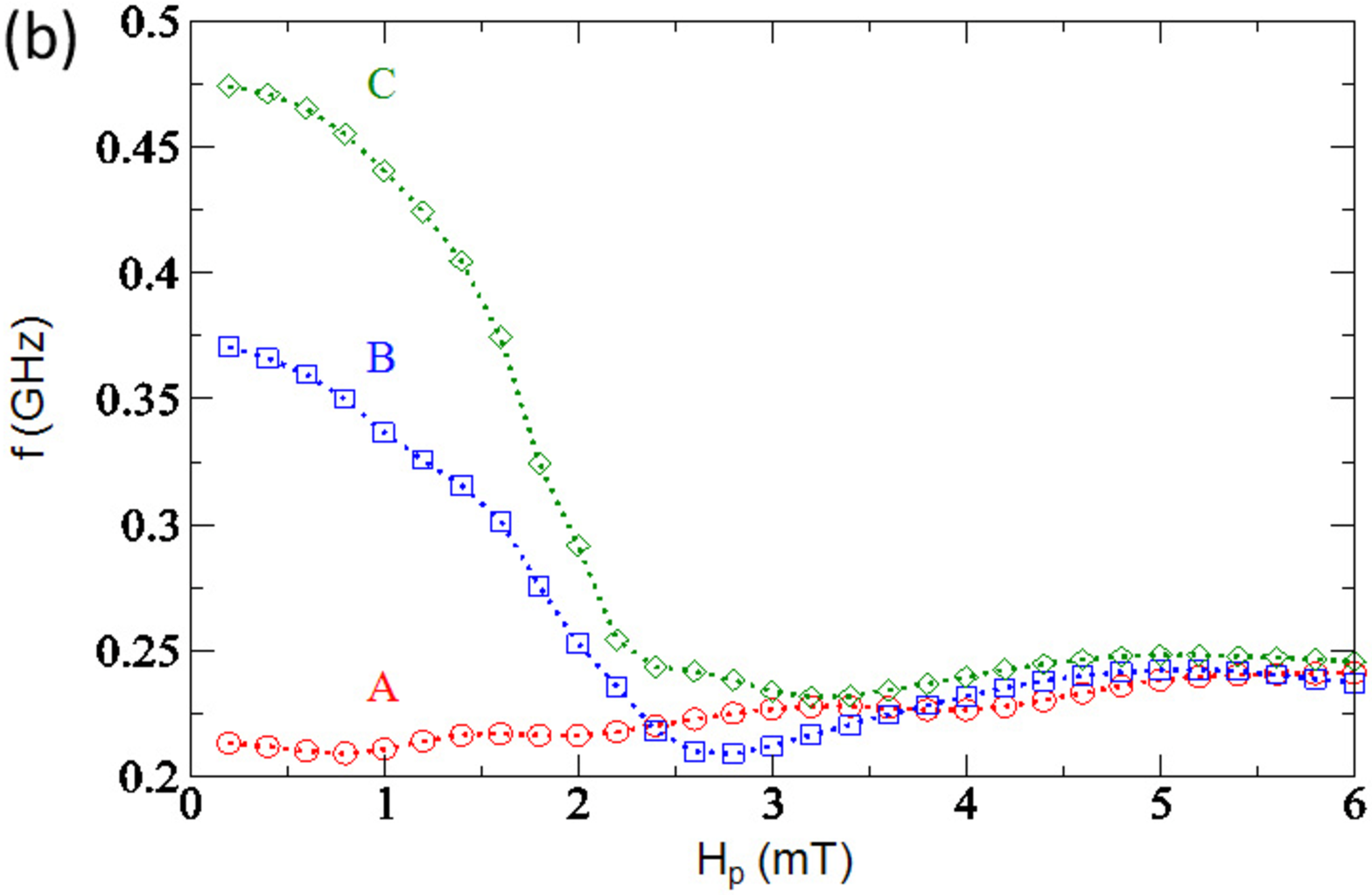}
\caption{(Color online). (a) Time evolution of gyration radius $r$ for
  the disordered sample in Fig.~\ref{fig:disorder_map} at points A, B and C
  for ${\cal D}=0.05$ with fixed $A$. To excite a vortex state,
  a Gaussian-type field pulse of 20~mT with 1~ns of the full width at
  half maximum along the $y$ direction is applied. The black line is
  for the case without disorder.  The small amplitude oscillations are
  due to the excitation procedure, which does not produce purely
  circular precession. (b) The gyration frequency averaged from $t=2$~ns
  to 12~ns as a function of field pulse amplitude at the points A, B
  and C.}
\label{fig:gyration_disorder}
\end{figure}

First, we study the dynamics of vortex gyration in the disordered
sample shown in Fig.~\ref{fig:disorder_map} for ${\cal D}=0.05$ with
fixed $A$. To move a vortex core to different positions, we apply
in-plane static magnetic fields to the sample. Maintaining the static
in-plane magnetic fields, we apply an additional Gaussian-type field
pulse of 20~mT along the $y$ direction with a 1~ns full width at half
maximum to excite the gyration motion.  This pulse is large enough to
induce a free vortex gyration with an initial radius that is much
larger than the disorder correlation length.  The gyration radius
decreases with time because of the energy dissipation through damping,
and eventually the vortex is trapped by the disorder potential when
its gyration radius is approximately the correlation length of the
disorder potential, which is 10~nm in the simulations.

Figure \ref{fig:gyration_disorder} shows the behavior of the gyration
for vortex center positions, A, B and C in
Fig.~\ref{fig:disorder_map}.  Position A is in a relatively flat
region of the disorder potential, position C is close
to a minimum in the disorder potential, and position B is intermediate
between the two.  Figure \ref{fig:gyration_disorder}(a) shows
the time evolution of the gyration radius $r$.  For ``free'' precession,
before the vortex becomes trapped, the gyration frequency and decay
rate are almost the same as the corresponding quantities in a disk
without disorder, and the vortex gyration motion is not changed
significantly by the disorder potential.  As we discuss in
Sec.~\ref{sec:discussion}, these results indicate that gyration in the
``free'' regime is minimally affected by disorder and there does not
appear to be an increase in the effective damping parameter in this
regime.  This result is in stark contrast to the behavior found for
the motion of a vortex domain wall through a disorder potential.  This
contrast will be discussed in more detail in
Sec.~\ref{sec:discussion}.  In the trapped regime, after approximately
90~ns, the decay rate of the precession radius increases dramatically
for positions B and C, those points close to minima in the disorder
potential.  This change in decay rate is associated with changes in
the precession rate.

Figure \ref{fig:gyration_disorder}(b) shows the change of the
gyration frequency as the field-pulse amplitude is varied, by
averaging over a fixed interval (from $t=2$~ns to 12~ns) after the pulse.
At small field-pulse amplitudes, the vortex
gyrates about its equilibrium position at points A, B and C with a frequency
characteristic of each pining site. At large pulse amplitudes, the
vortex core is depinned and gyrates with the frequency of the free
region. Between the two amplitudes, as a function of field pulse
amplitude, there is a crossover above which the gyration frequency
drops to the value of the free region due to depinning of the vortex core
from the pinning site.  Here, the transition is gradual because, for
some fields, the vortex core does not completely escape the pinning
potential and may even, for slightly higher fields, become repinned
during the averaging interval.  A related measurement in
Ref.~\onlinecite{compton2010} shows much more abrupt transitions in
Fig.~10(b).  However, there are significant differences in the
excitation pulse shape that strongly affect transition width as
illustrated in Fig.~10(d) in that same article.
Note that the crossover field here is a field range corresponding to
the vortex completely escaping the pinning potential without being
trapped again for a fixed time interval. 
For a Gaussian field pulse with
a 1~ns full width at half maximum, the crossover field for the time
average between 2~ns and 12~ns is approximately 2~mT for points B and
C. 

\section{Gyration in a single pinning potential}
\label{sec:VGsingle}

\subsection{Numerical results}

The existence of two regimes of motion of a vortex in a disordered
disk suggests that the behavior should be captured by a
simple model of a single pinning potential in the center of an
otherwise ideal disk.  Extreme examples of such samples, in which a
hole has been fabricated in the disk have been studied
experimentally\cite{uhlig2005,hoffmann2007} and theoretically.\cite{wysin2010}
Theoretical studies in
Ref.~\onlinecite{compton2010} are similar to ours but
focus on changes in the gyration frequency.
We choose the radius of the single pinning potential to be $r_{\rm pin}=$~10~nm
as a typical length scale of the potential. Inside the potential, the
magnetization increases quadratically from the center. We characterize
the potential depth as the ratio of reduced magnetization at the
center $\Delta M_{\rm c}$ to the saturation magnetization $M_{\rm s}$,
$\delta=\Delta M_{\rm c}/M_{\rm s}$. The vortex gyration is excited
by a Gaussian-type field pulse along
the $y$ direction of 20 
mT with a 1~ns full width at half maximum. When the strength of the
field pulse is large enough, the vortex core gyrates outside of the pinning
potential with a frequency that is determined by the geometry of the
disk. The radius of the gyration decreases due to the energy
dissipation through damping. When the vortex core enters the pinning
potential, it is trapped by the potential and the gyration frequency
changes.

\begin{figure}
\includegraphics[width=1\linewidth]{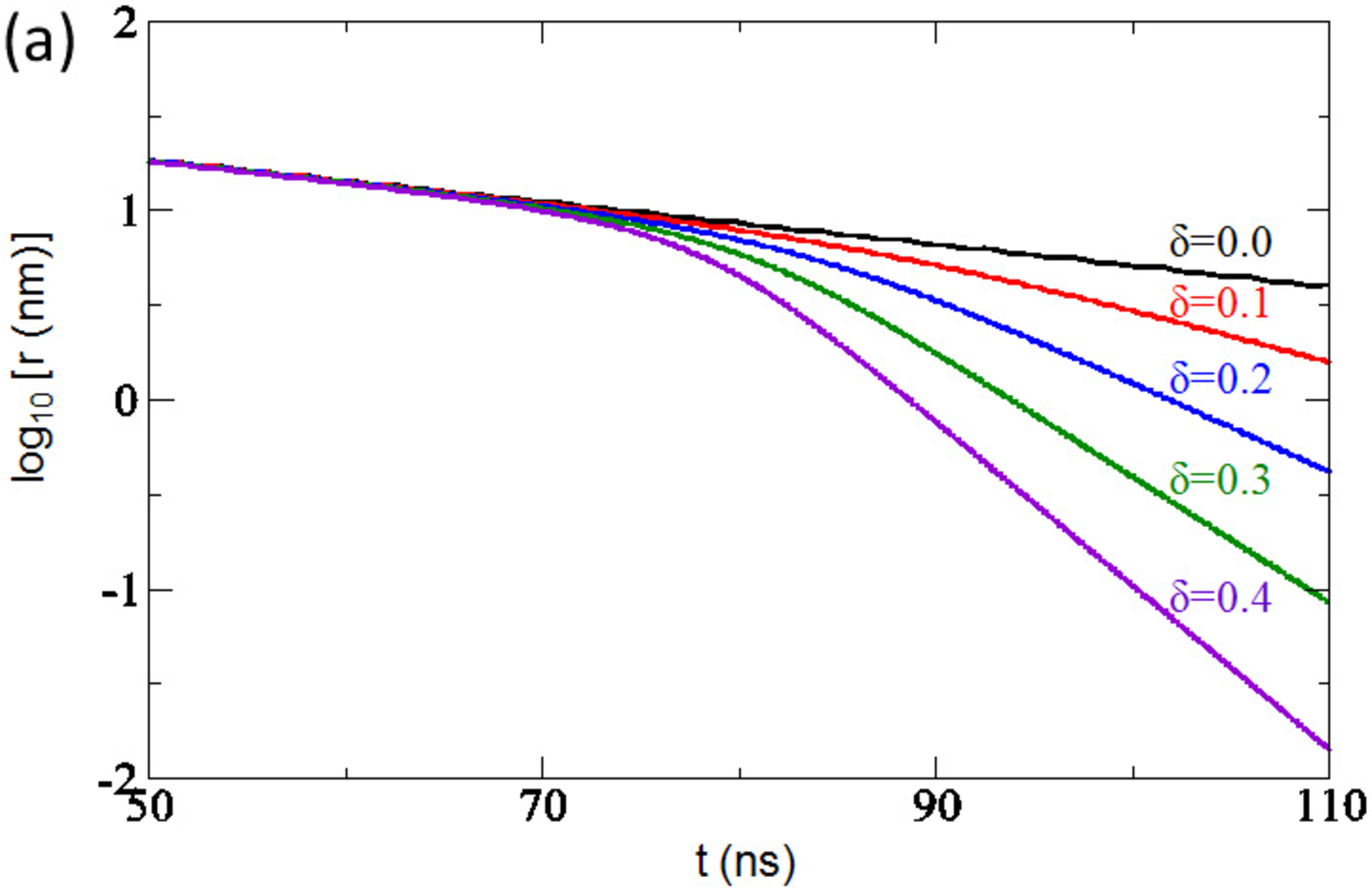}
\includegraphics[width=1\linewidth]{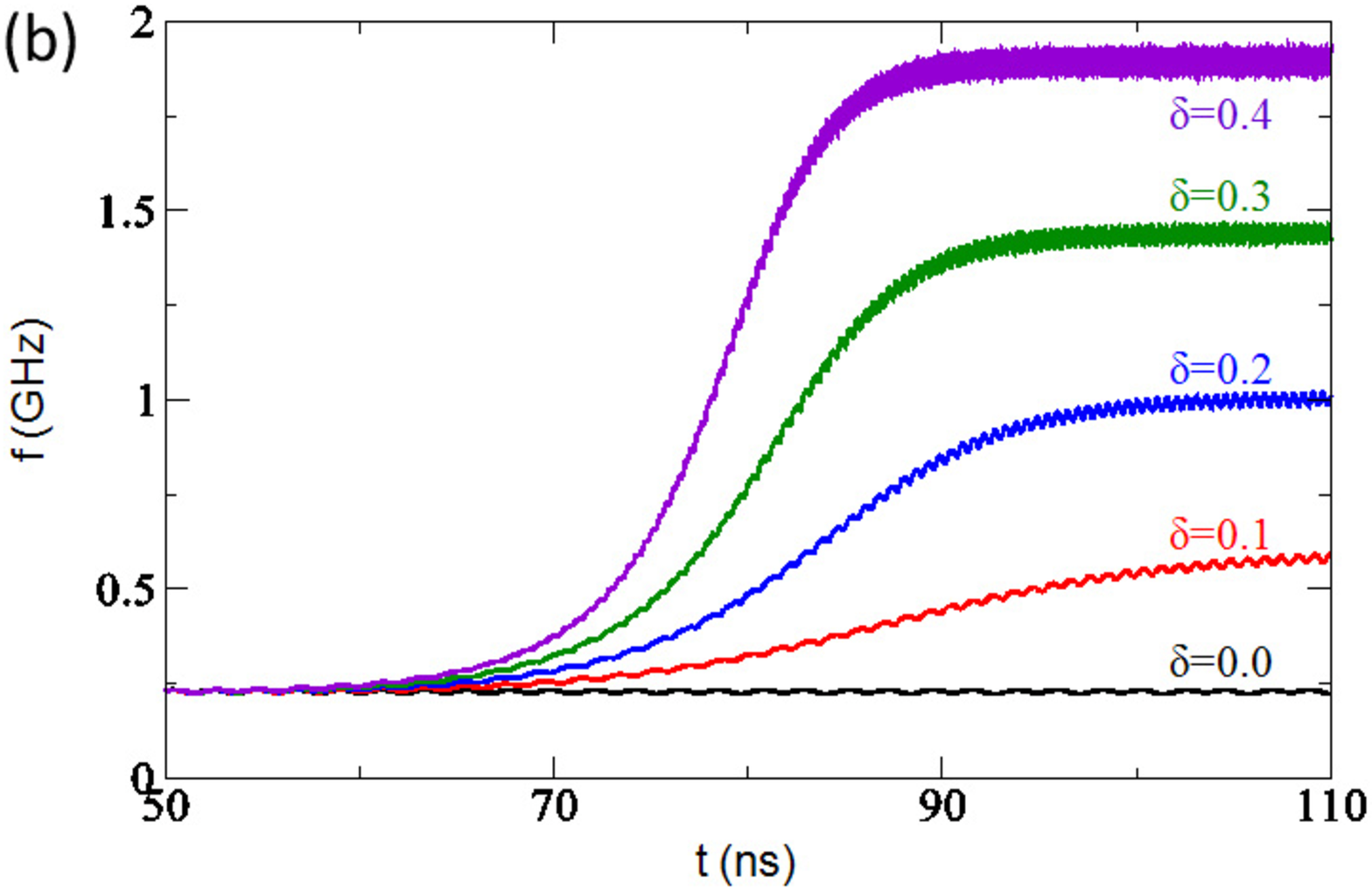}
\includegraphics[width=1\linewidth]{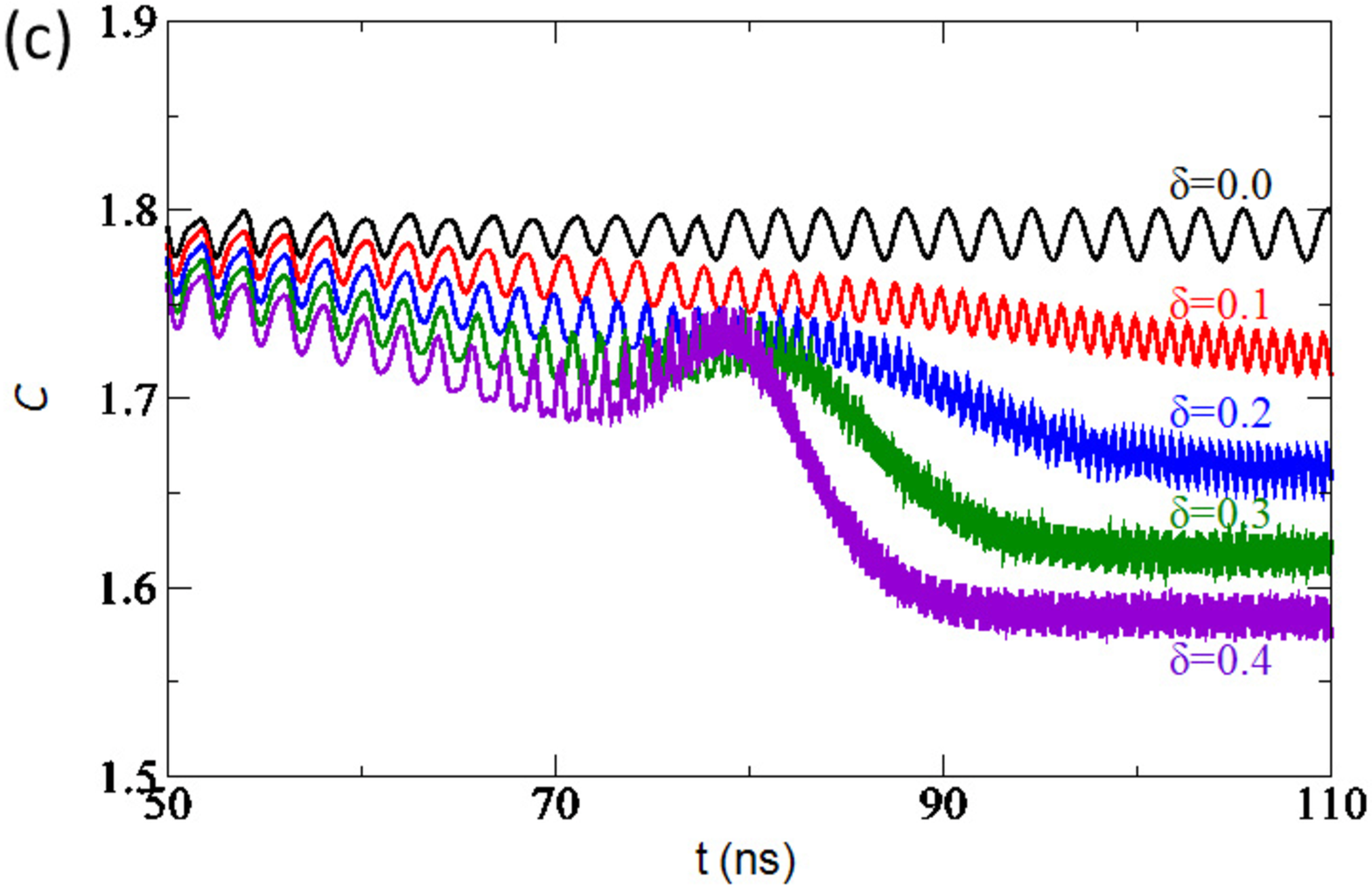}
\caption{(Color online). Time evolution of (a) gyration radius $r$,
  (b) gyration frequency $f$ and (c) deformation factor $C$ in a
  single pinning potential of the radius 10~nm for the depth ratio
  $\delta=0, 0.1, 0.2, 0.3$ and $0.4$ with fixed $A$. The gyration is
  excited by a 1~ns pulse of 20~mT.}
\label{fig:gyration}
\end{figure}

Figure \ref{fig:gyration} shows the time evolution of the gyration radius
$r$, the gyration frequency $f$, and the deformation factor $C$ for depth
ratios $\delta=0, 0.1, 0.2, 0.3$ and $0.4$ with fixed $A$. Here $f$ is
obtained from the angular velocity of gyration motion and $C$ is
calculated from Eq.~(\ref{eq:D_F_G}). From the slope of the logarithm
of $r$ we can obtain the decay time $\tau$. When the gyration radius
becomes close to 10~nm, which is the radius of the pinning potential,
$\tau$, $f$, and $C$ change, indicating that the vortex has become
trapped by the pinning potential. As $\delta$ increases, $\tau$ and
$C$ decrease while $f$ increases in the trapped regime.  Note that
additional oscillations superimposed on the curves have a frequency of
twice the gyration frequency and originate because the short pulse
with which we excite the gyration leads to a slightly elliptical orbit
for the vortex core.

\begin{figure}
\includegraphics[width=1\linewidth]{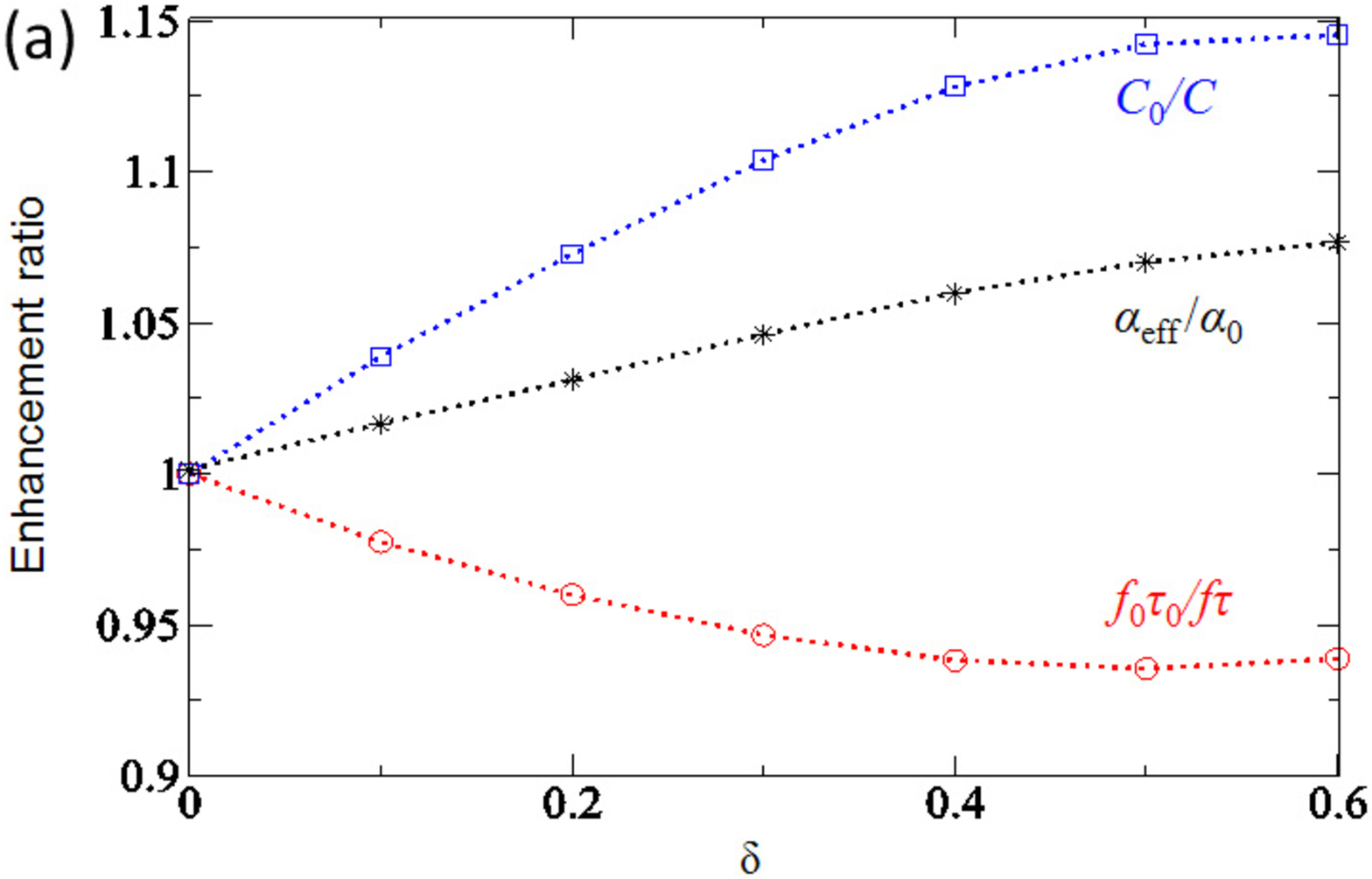}
\includegraphics[width=1\linewidth]{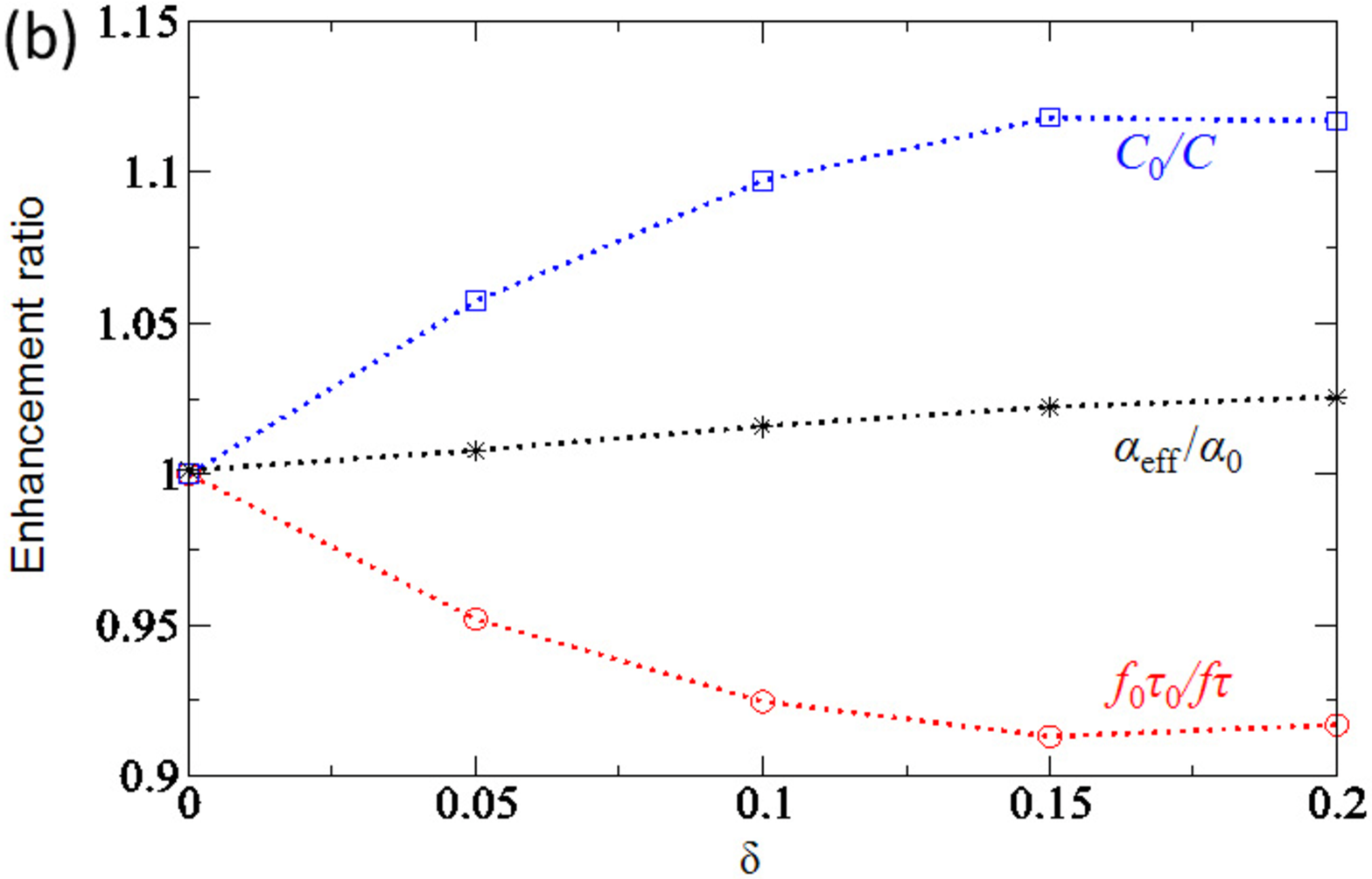}
\caption{(Color online). Enhancement ratio of $f\tau$, $C$, and
  $\alpha_{\rm eff}$ as a function of depth ratio $\delta$ for (a)
  fixed $A$ and (b) fixed $l_{ex}$. Subscript 0 indicates values in
  the free region before trapping.}
\label{fig:enhancement}
\end{figure}

Figure~\ref{fig:enhancement} shows the evolution of the different
factors in Eq.~(\ref{eq:damping_ratio}) as the depth of the pinning
potential is varied.  Since $f$ and $\tau$ can be
measured experimentally, for example, from the Kerr microscopy analysis, 
it is tempting to attribute the change in the decay to the change in
the effective damping.  However, Fig.~\ref{fig:enhancement} shows that,
in fact, most of the change is due to a change in the geometry of the
vortex through the factor $C$, which can be extracted from the
simulations.  In fact, ignoring the changes in $C$ 
would lead to the erroneous conclusion that damping decreases as the
depth of the well increases for the case considered here.  
The actual values of the enhancement rate
depend on various factors such as the type of disorder, geometry of
samples, and material properties such as the saturation magnetization
and the exchange constant.


\subsection{Radius dependence of the effective damping}

\begin{figure}
\includegraphics[width=1\linewidth]{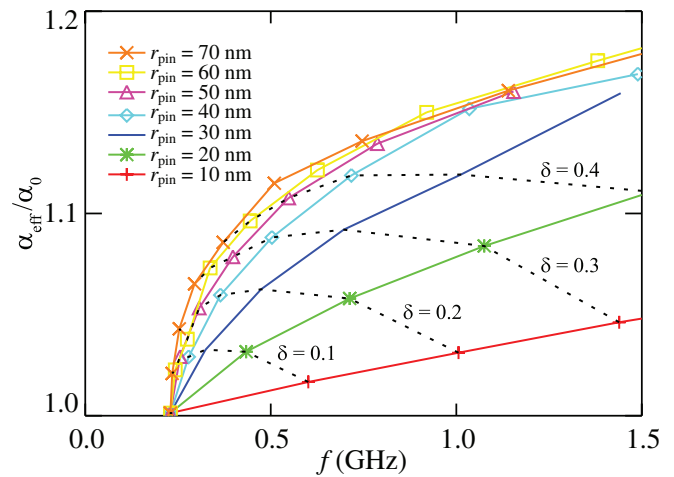}
\caption{(Color online). Frequency dependence of the effective damping
  for fixed $A$ and various pinning potential radii $r_{\rm pin}=10$~nm, 20~nm,
  $\ldots$ 70~nm
  and depth ratios $\delta=0, 0.1, 0.2, 0.3, 0.4$.  Solid lines
  indicate values for constant $r_{\rm pin}$ and dashed lines constant $\delta$. }
\label{fig:potential_radius}
\end{figure}

Figure \ref{fig:potential_radius} shows the frequency dependence of
the effective damping for the different radii of the pinning
potential $r_{\rm pin}$ with fixed $A$. As the frequency increases, the vortex
gets more excited increasing the effective damping. Note that
its slope increases as the radius of the pinning potential increases
and then saturates around $r_{\rm pin}\approx 60$~nm, which is approximately the
distance from the center of a disk to the point where the $z$ component
becomes zero, as shown in Fig.~\ref{fig:vortex}(b).

\section{Discussion}
\label{sec:discussion}

Disorder in magnetic samples can increase the energy dissipation rate
both for vortex gyration and domain wall propagation as shown
earlier.\cite{min2010}  An important contribution to this appears to
occur when the motion through the disorder excites additional degrees
of freedom in addition to the overall motion.  We refer to these
degrees of freedom as internal.  If these internal
degrees of freedom are not included explicitly in a collective
coordinate model,\cite{clarke2008} they will typically lead to an increase in the
effective damping parameter that describes the motion.  Vortex
gyration appears to give much smaller increase in the effective
damping than is found for domain wall motion\cite{min2010} but may be
more accessible experimentally.\cite{compton2006,compton2010}

The modeling results described in this article agree with the recent experiments and
theoretical analysis in Refs.~\onlinecite{compton2006} and
\onlinecite{compton2010}, demonstrating the frequency changes between
the free and trapped regimes.  Perhaps not
surprisingly, the transition between high-amplitude and low-amplitude
occurs when the gyration radius is comparable to the correlation
length. 
Here, we focus
not on the change of the frequencies between the two regimes, but on the
change of the effective damping constant, which is much smaller than
that of the frequency. 

Disorder has a negligible effect on the gyration decay rate when the
gyration amplitude is large.  However, the decay rate is increased by
disorder when gyration amplitude is small.  While a naive interpretation
would attribute changes in the product of gyration frequency and decay
time $f\tau$ to changes in the effective damping parameter
$\alpha$, a more detailed interpretation in
the context of the collective coordinate approach shows that the
majority of the change in $f\tau$ is due to disorder-induced changes
in the deformation factor $C$, and that the change in $\alpha$ is modest.  
Even though the change in the frequency can be
successfully modeled assuming the  deformation factor $C$ is
constant, evaluating the effective damping correctly requires detailed
calculation. 


Comparing these vortex gyration results with the results of similar
calculations on vortex wall propagation, it is clear that the domain
wall motion is more sensitive to disorder.
For field or current
induced vortex wall propagation in the presence of $M_{\rm s}$
fluctuations with fixed $A$,\cite{min2010} the effective damping is
enhanced almost two times for ${\cal D}=0.05$ (5 \% average
fluctuation of $M_{\rm s}$). For vortex gyration in a single pinning
potential with fixed $A$, however, the damping is enhanced by only
6~\% even for $\delta=0.4$ (40 \% of reduced magnetization at the 
center), as shown in Fig.~\ref{fig:enhancement}.

One possible reason for this difference is the appearance of half
antivortices in domain wall propagation. As the vortex wall propagates
either by field or current along a magnetic nanowire, energy
dissipates mostly through the motion of the vortex core and two half
antivortices.\cite{min2010} The relative motions of these structures
are examples of internal 
degrees of freedom.  When these are excited by moving in the disorder potential, the
energy dissipation rate increases.  In a disk, on the other hand, there are no
antivortices, as seen in Fig.~\ref{fig:vortex}.  When vortices are
driven to large amplitudes, antivortices can appear and can lead to
core reversal\cite{guslienko2008} when the system is driven hard
enough.  In the simulations we consider here, we are not in this
regime and the distortions of the vortex are relatively small. The
lack of antivortices in these simulations is consistent with a reduced
excitation of internal degrees of freedom as compared to the vortex
wall propagation, resulting in the smaller enhanced effective damping.

In summary, we have demonstrated that disorder enhances the effective
damping, and the enhancement ratio can be estimated up to the
deformation factor by a vortex gyration experiment in a magnetic disk
with a single pinning potential at the center. By measuring the
frequency $f$ and decay time $\tau$ in free and trapped regions, we
can estimate the enhancement ratio of the effective damping times the
deformation factor. 

\acknowledgments 
The work has been supported in part by the
NIST-CNST/UMD-NanoCenter Cooperative Agreement.  We thank Te-Yu Chen
and Paul Crowell for useful discussions about the experiments.


\appendix*

\section{Fixed $A$ vs fixed $l_{\rm ex}$}

In this article we are using a simple approximation to model the effect
of thickness fluctuations.  Since the important energies in the
problem are the magnetostatic (stray field) energy, the Zeeman energy, and the
exchange energy, we can capture the changes in the first two by
locally varying the saturation magnetization $M_{\rm s}$.
In the main body of the text, we report results for vortex gyration as
we model thickness variations by changing $M_{\rm s}$ while keeping
the exchange constant $A$ constant.  An alternate approach would be to
keep the exchange length $l_{\rm ex}=[2A/(\mu_0 M_{\rm s}^2)]^{1/2}$
constant.  In this appendix, we describe the quantitative but not
qualitative changes that result.  

For a fixed $l_{\rm ex}$, a disorder
value of ${\cal D}=0.0125$ gives a similar variation in gyration
frequency as a value of ${\cal D}=0.05$ for constant $A$.  
We attribute the higher sensitivity to disorder with fixed $l_{\rm
  ex}$ to changes in exchange energy in the region outside the vortex
core.  Briefly, Eq.~(\ref{eq:lexDef}) shows that $A$ becomes a function of
$M_{\rm s}$ when $l_{\rm ex}$ is fixed. Consequently, the exchange
energy associated with the curling of the magnetization around the
vortex core decreases when $M_{\rm s}$ decreases.  

For a single pinning potential, simulations for the fixed $l_{\rm ex}$ case show similar
trends as seen in Fig.~\ref{fig:gyration} but with larger frequency
and smaller decay time compared with 
the fixed $A$ case.  In addition, we see in Fig.~\ref{fig:enhancement}(b)
that the fixed $l_{\rm ex}$ case leads to a smaller change in the
effective damping for a given change in $f\tau$ as compared to the
fixed $A$ case.

To
understand the difference between $M_{\rm s}$ fluctuations with fixed
$A$ and those with fixed $l_{\rm ex}$, consider a simple model of a
vortex in a thin film disk with thickness $z$.  We
estimate the core energy and how it varies with the reduction in
the magnetization in the pinning potential for both fixed $A$ and
fixed $l_{\rm ex}$.
We model the pinning
potential as a circular region of radius $r_{\rm pin}$ of uniformly
reduced magnetization $M_{\rm pin}$. 
We assume that the vortex can be described as a core region within a
radius $r_{\rm c}$ where the magnetization points out of plane, and an outer
region $r > r_{\rm c}$ where the magnetization is directed azimuthally. Our
approach is to estimate the change in vortex energy as $M_{\rm pin}$ is
varied in the pinning potential region , $r < r_{\rm pin}$.  For
simplicity, we just consider the case for $r_{\rm c}<r_{\rm pin}$.

In the core region, the magnetization is uniform, and the energy density
is magnetostatic, thus we can approximate the energy of the core as
\begin{equation}
E_{\rm core}= \pi r_{\rm c}^2 z \frac{1}{2} \mu_0 M_{\rm pin}^2.
\end{equation}
In this expression we have assumed that $z \ll r_{\rm c}$ for simplicity.

The divergence of the magnetization in the outside region, $r > r_c$,
is zero and the 
magnetization is perpendicular to the stray fields from the core
region, so the energy in the outside region is entirely exchange
energy. 
Since the magnetization outside $r_{\rm pin}$ does not depend on
$M_{\rm pin}$ and the 
geometry of the vortex is fixed (exactly in this simple model and
approximately in a full simulation), the energy of the magnetization
from $r_{\rm pin}$ out to $R$, the radius of the disk, does not
change. Ignoring that constant contribution leaves the
$M_{\rm pin}$-dependent change in the exchange energy from $r_{\rm c}$ to $r_{\rm pin}$
\begin{eqnarray}
E_{\rm outside} & = &\int dv \frac{A}{M_{\rm pin}^2} \left(\frac{d{\bf M}}{d{\bf x}}\right)^2 \nonumber \\
& = & 2\pi z A \ln(r_{\rm pin}/r_{\rm c}).
\end{eqnarray}
As a further approximation, we ignore any exchange energy associated
with the sharp transition from vertical to in-plane magnetization at
the core boundary, $r_{\rm c}$.

So far, the core radius has been left as a variable, and we determine
its value by minimizing the total energy $E_{\rm total}=E_{\rm core}
+E_{\rm outside}$.  The result is
\begin{equation}
r_{{\rm c, min}} =\sqrt{\frac{2A}{\mu_0 M_{\rm pin}^2}} \equiv l_{\rm ex},
\label{eq:lex}
\end{equation}
and the minimized energy is
\begin{equation}
\label{eq:E_A}
E_{\rm total} = \pi z A \left[1+\ln\left(\frac{r_{\rm pin}^2\mu_0M_{\rm pin}^2}{2A}\right)\right]
\end{equation}
or equivalently,
\begin{equation}
\label{eq:E_lex}
E_{\rm total} = \pi l_{\rm ex}^2 z \frac{1}{2}\mu_0 M_{\rm pin}^2
\left[1+2\ln\left(\frac{r_{\rm pin}}{l_{\rm ex}}\right)\right].
\end{equation}

Therefore, from Eqs.~(\ref{eq:E_A}) and (\ref{eq:E_lex}), the ratio of
the total energy changes using constant $A$ compared to constant
$l_{\rm ex}$ is
\begin{equation}
\frac{\left.\frac{d E_{\rm total}}{d M_{\rm pin}}\right|_{l_{\rm ex}}}
     {\left.\frac{d E_{\rm total}}{d M_{\rm pin}}\right|_A}  
= 1+2\ln\left(\frac{r_{\rm pin}}{l_{\rm ex}}\right).
\end{equation}
For $r_{\rm pin}$ = 10~nm and $l_{\rm ex}$ = 5.7~nm, the ratio is
$\approx$ 2.1.  That is, the energy is more sensitive to variations in
$M_{\rm pin}$ for fixed $l_{\rm ex}$ than for fixed $A$.



With $A$ fixed, a decrease in $M_{\rm pin}$ results in an expansion of
the core radius [see Eq.~(\ref{eq:lex})] such that the decreased magnetostatic energy
density within the core is compensated by an increased core volume,
yielding no net change in the magnetostatic energy.  The net energy
change is due to a decrease in the exchange energy as the region $r_{\rm c} < r <
r_{\rm pin}$ becomes smaller as $r_{\rm c}$ increases.

In contrast, for fixed $l_{\rm ex}$, the geometry of the core is
fixed, and there is a decrease in the magnetostatic energy of the core
associated with a decrease in $M_{\rm pin}$.  Further, with $l_{\rm ex}$ fixed
it can be seen from (\ref{eq:lex}) that $A$ must decrease with $M_{\rm
  pin}$, and this results in reduced exchange energy calculated in
the region $r_{\rm c} < r < r_{\rm pin}$.


\end{document}